\documentstyle[twocolumn,aps]{revtex}

\input epsf
\epsfverbosetrue

\begin{document}

\title{Large-scale magnetic fields from
hydromagnetic turbulence in the very early universe}
\author{Axel Brandenburg\thanks{Permanent address:
Department of Mathematics and Statistics,
University of Newcastle upon Tyne, NE1 7RU, UK.
Electronic address: Axel.Brandenburg@Newcastle.ac.uk}}
\address{Nordita, Blegdamsvej 17, DK-2100 Copenhagen \O, Denmark}
\author{Kari Enqvist\thanks{Electronic address: enqvist@pcu.helsinki.fi}}
\address{Department of Physics, P.O. Box 9, FIN-00014 University of Helsinki, Finland}
\author{Poul Olesen\thanks{Electronic address: polesen@nbivax.nbi.dk}}
\address{The Niels Bohr Institute, Blegdamsvej 17, DK-2100 Copenhagen \O, Denmark}

\maketitle

\begin{abstract}
We investigate hydromagnetic turbulence
of primordial magnetic fields using  magnetohydrodynamics (MHD)
in an expanding universe. We present the basic, covariant
MHD equations, find solutions for MHD waves in the early universe, and
investigate the equations numerically for random magnetic fields
in two spatial dimensions. We find the formation of magnetic structures
at larger and larger scales as time goes on.
In three dimensions we use a cascade (shell) model, that has been rather
successful in the study of certain aspects of hydrodynamic turbulence.
Using such a model we find that after ${\cal O}(10^9)$ times the initial time
the scale of the magnetic field fluctuation (in the comoving frame) has
increased by 4-5 orders of magnitude as a consequence of an inverse
cascade effect (i.e. transfer of energy from smaller to larger scales).
Thus {\it at large scales} primordial magnetic fields are considerably
stronger than expected from considerations which do not take into
account the effects of MHD turbulence.
\end{abstract}

\section{Introduction}
It has been suggested 
that primordial magnetic fields might arise during the early cosmic phase
transitions \cite{pt}, and recently it has been shown that magnetic fields
are indeed a stable feature of  the electroweak 
phase transition \cite{davies}. In a first order phase transition magnetic
fields could also be generated when bubbles of the new vacuum collide, whence
a ring of magnetic field may arise in the intersecting region \cite{kibble}.
It has also been suggested that at very high temperatures the ground state
of a non-abelian particle theory is a ``ferromagnetic''
vacuum with a permanent non-zero magnetic field \cite{us}. 
The plasma of the early universe has a high conductivity so that a
primordial magnetic field would be imprinted on the comoving plasma and
would dissipate very slowly \cite{dissip}. Such a field could 
then contribute to the seed field needed to understand the presently observed 
galactic magnetic fields \cite{fields}, which have been measured both in the
Milky Way and in other spiral galaxies, including their halos. Typically the
observed present day magnetic field is of the order of ${\cal O}(10^{-6})$ G. 

Locally the primordial field could be very large; it is limited only by
the magnetic energy density contribution to
 primordial nucleosynthesis, which constraints the field at the time of 
nucleosynthesis by
$\rho_B=B_{\rm rms}^2/8\pi\le 0.3\rho_\nu$, where $\rho_\nu$ is the energy density
of 3 species of massless neutrinos
\cite{dario}.
Because flux conservation implies that $B_{\rm rms}\sim R^{-2}$, where 
$R$ is the scale factor, the field could have been much stronger at earlier
times. On dimensional grounds, a typical value of the magnetic field fluctuation
should be $B_{\rm rms}\sim T^2$, so that at the time of the electroweak phase
transition one could locally obtain fields as high as $10^{24}\,$G.
Depending on how such a strong, random magnetic field scales at large
distances, it could \cite{poul} be the seed field needed to explain the
magnetic fields observed on the scale of galaxies and larger.
                                        
However, even assuming that a primordial magnetic field is created at some very
early epoch, a number of issues remain to be worked out before
one can say anything definite about the role of primordial fields
in generating galactic magnetic fields. At earliest times magnetic fields
are generated by particle physics processes with length scales typical to
particle physics. If the inflation hypothesis proves correct, then
after inflation rather long correlation lengths are possible \cite{ven}.
The question is if it is at all possible
for the small scale fluctuations to grow to large scales, and what exactly is
the scaling behaviour of $B_{\rm rms}$ or the correlator
$\langle B(r+x) B(x)\rangle$. Even in an inflationary scenario it would
be of interest to see if the relatively large scale can grow even further. To
study these problems one needs to consider the detailed evolution of the
magnetic field  to account for such issues as what happens when uncorrelated
field regions come into contact with each other during the course of the
expansion of the universe. In general, turbulence is an essential
feature of such phenomena. These questions can only be answered by considering
magnetohydrodynamics (MHD) in an expanding universe.\cite{frozenin}
It is the main purpose of this paper to
investigate the subsequent development of the primordial field. Expressed in a
general way, our conclusion turns out to be that MHD turbulence
is operative, and hence the scale of magnetic fields is considerably larger than
one would expect if MHD turbulence was ignored. This means that the previous
estimates of the strength of the primordial magnetic field ``today" need to be
reconsidered.

We begin by posing the basic equations and consider certain simplified models.
A full 1+3-dimensional numerical simulation would be desirable, but
is beyond the scope of the present paper. In Sec.~II we derive the
relativistic MHD equations for relativistic plasma, which is appropriate for
the very early universe. In Sec.~III we discuss the appearance of waves in
relativistic MHD. To illucidate the various MHD effects pertaining the early
universe, we also present numerical solutions to the MHD equations for
a two-dimensional slice. In Sec.~IV we study a cascade model, that
reflects important properties of fully three-dimensional turbulence.
The cascade model has been rather successful in ordinary hydrodynamics.
We find that in the early universe magnetic energy is
transferred from small scales to large scales. We also compute
the correlation function $\langle B(r+x) B(x)\rangle$ in the
cascade model. In Sec.~V we offer an interpretation of our results.
\section{Relativistic MHD in the expanding universe}
We begin by presenting a derivation of the fully general relativistic
MHD equations (see also ref.\cite{relMHD}, where further references can be
found), which we rewrite in a form suitable for our numerical work.
We consider the early universe as consisting of ideal fluid with
an equation of state of the form $p=\frac{1}{3}\rho$, where $p$ is pressure,
$\rho$ the energy density, and the speed of light is set to unity. We further
assume that the fluid supports a (random) magnetic field.
The energy-momentum tensor is then given by
\begin{eqnarray}
T^{\mu\nu}&=&(p+\rho) U^\mu U^\nu + p g^{\mu\nu} \nonumber \\
&+&{1\over4\pi}\left(F^{\mu\sigma} {F^\nu}_\sigma
-{1\over4}g^{\mu\nu}F_{\lambda\sigma}F^{\lambda\sigma}\right),
\label{energymomentum}
\end{eqnarray}
and $U^\mu$ is the four-velocity of the plasma,
normalized as $U^\mu U_\mu=-1$,  and $F_{\mu\nu}=
\partial_\mu A_\nu-\partial_\nu A_\mu$ is the electromagnetic field tensor.
Note that, as long as diffusion can be neglected, the presence of the magnetic
field does not change the equation of state. 

The magnetic energy is assumed to be much smaller than the radiation
energy, so that it can be neglected as far as the expansion of the
universe is concerned. We therefore assume a flat, isotropic and homogeneous 
universe with a Robertson-Walker metric
$ds^2=-dt^2+R^2(t) d{\bf x}^2$. Although the magnetic field generates local
bulk motion, this may still be consistent with isotropy and homogeneity at
sufficiently large scales, e.g.\ if the magnetic field is random,
i.e.\ statistically homogeneous and isotropic on scales much larger than the
intrinsic correlation scale of the field. Even very large magnetic fields,
together with the ensuing very fast bulk motion, might not contradict isotropy
and homogeneity.
The equations of motion for the fluid arise from energy-momentum conservation 
\begin{equation}
{T^{\mu\nu}}_{;\nu}\equiv{1\over\sqrt{-g}}{\partial\over\partial x^\nu}
\sqrt{-g}{T^{\mu\nu}}+\Gamma^\mu_{\nu\lambda}{T^{\nu\lambda}}=0.
\label{tmunu}
\end{equation}
The Maxwell equations read
\begin{equation}
{F^{\mu\nu}}_{;\nu}=4\pi J^\mu,\quad
F_{[\mu\nu,\lambda]}=0.
\label{maxwell}
\end{equation}
We define $F_{\mu\nu}$ in terms of the electric and magnetic fields
\begin{equation}
F_{i0}=RE^i,\quad F_{ij}=\epsilon_{ijk}R^2B^k,
\label{ejab}
\end{equation}
where latin letters go from 1 to 3. With this definition the expression for
the total energy has no $R$-factors and takes therefore the familiar form
\begin{equation}
T^{00}=(p+\rho)\gamma^2 - p +{1\over8\pi}({\bf B}^2+{\bf E}^2),
\label{T00}
\end{equation}
where $\gamma=U^0$. 

In order to solve (\ref{tmunu}) and (\ref{maxwell}) we rewrite the
equations of motion explicitly in 3+1 dimensions. We start by writing
(\ref{tmunu}) as
\begin{eqnarray}
{1\over\sqrt{g}}{\partial\over\partial x^\nu}
\left[\sqrt{g}(p+\rho) U^\mu U^\nu\right]
&+&\Gamma^\mu_{\sigma\lambda}(p+\rho) U^\sigma U^\lambda
\nonumber \\
&+&g^{\mu\nu}{\partial p\over\partial x^\nu}
=F^{\mu\nu} g_{\nu\sigma} J^\sigma,
\label{alku}
\end{eqnarray}
where $\sqrt{-g}=R^3$, and the nonvanishing Christoffel symbols are
$\Gamma^0_{ij}=R\dot{R}\delta_{ij}$ and
$\Gamma^i_{0j}=(\dot{R}/R)\delta^i_j=\Gamma^i_{j0}$.
It is useful to define $U^i=\gamma R^{-1} v^i$, because then
the normalization $U^\mu U_\mu=-1$ gives the familiar form
for the Lorentz factor $\gamma=(1-{\bf v}^2)^{-1/2}$.

For $\mu=i$ we obtain
\begin{eqnarray}
{\partial R^4{\bf S}\over\partial t}={1\over R}[
&-&(\mbox{\boldmath $\nabla$}\cdot{\bf v})(R^4 {\bf S})
-({\bf v}\cdot\mbox{\boldmath $\nabla$})(R^4 {\bf S}) \nonumber \\
&-&\mbox{\boldmath $\nabla$} (R^4 p) +(R^3{\bf J})\times(R^2{\bf B})],
\label{Sexpression2}
\end{eqnarray}
where ${\bf S}=(p+\rho)\gamma^2{\bf v}$. It should be noticed that in
this equation all quantities are scaled by the appropriate powers of
$R$. Thus, e.g.  $R^4{\bf S}$ is expected to be independent of $R$,
because $p+\rho$ scales like 1/$R^4$, and ${\bf v}$ is expected to be
independent of $R$. Also, $\mbox{\boldmath $\nabla$}$ occurs always
multiplied by 1/$R$, or, alternatively, the operator $\partial/\partial t$
is replaced by itself multiplied by $R$, which means that
time is replaced by conformal time $\tilde{t}=\int dt$/$R$. To
emphasize this, it is convenient to introduce new scaled
``tilde"-variables,
\begin{eqnarray}
&\tilde{{\bf S}}=R^4{\bf S}&,\quad\tilde{p}=R^4p,\quad\tilde{\rho}=R^4\rho,
\quad\tilde{{\bf B}}=R^2{\bf B},\nonumber \\
&\tilde{\bf J}=R^3{\bf J}&,\quad{\rm and}\quad\tilde{\bf E}=R^2{\bf E}.
\label{tlde}
\end{eqnarray}
It should be noticed that ${\bf v}$ is not scaled.
Equation (\ref{Sexpression2}) can then be written
\begin{equation}
{\partial\tilde{{\bf S}}\over\partial \tilde{t}}=
-(\mbox{\boldmath $\nabla$}\cdot{\bf v})\tilde{{\bf S}}
-({\bf v}\cdot\mbox{\boldmath $\nabla$})\tilde{{\bf S}}
-\mbox{\boldmath $\nabla$}\tilde{p}+\tilde{\bf J}\times\tilde{{\bf B}}.
\label{tld2}
\end{equation}
For $\mu=0$ we obtain, using scaled quantities,
\begin{eqnarray}
\left(1-{1\over4\gamma^2}\right){\partial\ln\tilde\rho\over\partial\tilde t}
&+&{\partial\ln\gamma^2\over\partial\tilde t}
+{\bf v}\cdot\mbox{\boldmath $\nabla$}\ln(\tilde\rho\gamma^2)
\nonumber \\
&+&\mbox{\boldmath $\nabla$}\cdot{\bf v}
={\tilde{\bf J}\cdot\tilde{\bf E}\over(\tilde p+\tilde\rho)\gamma^2}.
\label{dlnrhodt1}
\end{eqnarray}
In order to solve this equation numerically with an explicit code we need
to eliminate the time derivative $\partial\ln\gamma^2/\partial t$.
To this end we first solve the normalization condition for $\gamma^2$,
\begin{eqnarray}
\gamma^2={1\over2}+\left({1\over4}
+{\tilde{\bf S}^2\over(\tilde p+\tilde\rho)^2}\right)^{1/2},
\label{gamma2}
\end{eqnarray}
where we have used $\rho+p=\frac{4}{3}\rho$ for later convenience.
We then differentiate
\begin{eqnarray}
{\partial\ln\gamma^2\over\partial\tilde t}={1\over\gamma^2(2\gamma^2-1)}
{\partial\tilde{\bf S}^2/\partial\tilde t\over(\tilde p+\tilde\rho)^2}
-{\gamma^2-1\over\gamma^2-\frac12}{\partial\ln\tilde\rho\over\partial\tilde t}.
\label{dgamma2dt}
\end{eqnarray}
Combining (\ref{dlnrhodt1}) and (\ref{dgamma2dt}) we obtain a final equation
suitable for numerical work
\begin{eqnarray}
{2\gamma^2+1\over 4\gamma^2(2\gamma^2-1)}{\partial\ln\tilde{\rho}\over
\partial\tilde{t}}
&=&-{\partial\tilde{{\bf S}}^2/\partial\tilde{t}\over\left({4\over3}\tilde{\rho}
\gamma\right)^2 (2\gamma^2-1)} \nonumber \\
&-&{\bf v}\cdot\mbox{\boldmath $\nabla$}
\ln(\tilde{\rho}\gamma^2)-\mbox{\boldmath $\nabla$}\cdot{\bf v}
+{\tilde{\bf J}\cdot\tilde{\bf E}\over{4\over3}\tilde{\rho}\gamma^2}.
\label{rhoexpression2}
\end{eqnarray}
In this equation only scale invariant quantities enter.

The Maxwell equations can be written explicitly as
\begin{equation}
{\partial \tilde{{\bf B}}\over\partial\tilde{t}}=
-\nabla\times\tilde{\bf E},\quad\mbox{\boldmath $\nabla$}\cdot\tilde{{\bf B}}=0,
\label{induct1}
\end{equation}
and
\begin{equation}
\tilde{\bf J}={1\over4\pi }\mbox{\boldmath $\nabla$}\times\tilde{{\bf B}}
-{\partial \tilde{\bf E}\over\partial \tilde{t}},
\quad\mbox{\boldmath $\nabla$}\cdot\tilde{\bf E}=4\pi\tilde\rho_e
\end{equation}
where $\rho_e$ is the charge density and $\tilde\rho_e=R^3\rho_e$. Further,
\begin{equation}
\tilde{\bf E}=-{\bf v}\times\tilde{{\bf B}},
\label{vxB}
\end{equation}
which is valid in the limit of high conductivity \cite{relMHD}.
Again, these equations have the natural scaling properties with respect to
powers of $R$.
We emphasize that in the relativistic regime the displacement current,
$-\partial\tilde{\bf E}/\partial t$, cannot be neglected.
However, in all cases considered we were able to solve for
$-\dot{\tilde{\bf E}}=\dot{{\bf v}}\times\tilde{\bf B}
+{\bf v}\times\dot{\tilde{\bf B}}$
iteratively by evaluating $\dot{{\bf v}}$ and $\dot{\tilde{\bf B}}$ from the
previous iteration.

The equation of energy conservation is ${T^{0\nu}}_{;\nu}=0$, or
\begin{equation}
{1\over R^3}{\partial\over\partial t}\left(R^3 T^{00}\right)
+{\partial\over\partial x^j}T^{0j}+R\dot{R}T^{jj}=0,
\end{equation}
but since ${T^\nu}_\nu=0$, we have
$T^{jj}={T^j}_j/R^2=-{T^0}_0/R^2=T^{00}/R^2$
and therefore the energy equation is
\begin{equation}
{\partial\over\partial t}R^4 T^{00}=
-{\partial\over\partial x^j}R^4 T^{0j},
\end{equation}
or integrated over the whole space
\begin{equation}
{d R^4 E_{\rm tot}\over dt}=0,
\end{equation}
where
\begin{equation}
E_{\rm tot}=\int T^{00}d^3x\equiv\left\langle T^{00}\right\rangle.
\end{equation}
Hence $R^4 E_{\rm tot}$ is conserved.

The conclusion from the above expressions is thus that {\it the MHD
equations in an expanding universe with zero curvature are the same as
the relativistic MHD equations in a non-expanding universe, provided the
dynamical quantities are replaced by the scaled ``tilde" variables, and provided
conformal time $\tilde{t}$ is used.} The effect of this is, as usual, that the
expansion slows down the rate of dynamical evolution.

It should be noted that the velocity ${\bf v}$ is the bulk velocity.
Thus, in general we expect that ${\bf v}$ is nonrelativistic.
This is  physically reasonable since, although the gas particles move
with velocity near unity, we expect no strong collective effects which could 
give rise to a relativistic bulk velocity.
The equations for nonrelativistic bulk motions of a relativistic gas
are given in Appendix A.

In the early universe conductivity is large, and hence the diffusion length is
also large. The conductivity of the isotropic relativistic electron
gas, which interacts with heavy (non-relativistic) ions, is related to
the Coulomb scattering cross section and reads \cite{con}
\begin{equation}
\sigma={\omega_p^2\over 4\pi\sigma_{\rm coll}n_e}\simeq {T\over 3\pi\alpha},
\label{cond}
\end{equation}
where $\omega_p$ is the plasma frequency, $\sigma_{\rm coll}$ is the
collision cross section and $\alpha$ is the fine structure constant. 
This result is valid for fields smaller than
the critical field $B_c=m_e^2/e=4.41\times 10^{13}\,$G, above which
the electrons cannot be treated as free, and the conductivity (\ref{cond})
should be multiplied by a factor $B/B_c$. On dimensional grounds,
conductivity of the fully relativistic Standard Model gas will also
scale as $\sigma\sim T$. 
The expansion rate of the radiation dominated universe is given by
\begin{equation}
H\equiv \frac{\dot R}R = {1\over 2t}=\sqrt{8\pi^3 g_*\over 90}{T^2\over M_P}~,
\label{hubble}
\end{equation}
where $g_*$ is the number of the effective degrees of freedom, and
$M_P=1.2\times 10^{19}$ GeV is the Planck mass. Equation (\ref{hubble}) also
provides the time-temperature relationship \cite{kolb}, and the
inverse the length scale of the universe.
A measure of the importance of the diffusion is the magnetic Reynolds number,
which may be defined as $Re=Lv\sigma$, where $L$ and $v$ are
respectively the typical length scale and velocity in the system under 
consideration. A Reynolds number less than one means that diffusion dominates.
In the early
universe, say at the electroweak phase transition $T_{\rm EW}\simeq
{\cal O}(100)$ GeV where in the Standard Model $g_*=106.75$,
the Reynolds number is huge; typically
\begin{equation}
Re_{\rm U}\sim v\sigma H^{-1}\sim {M_P\over T}\sim 10^{17}~,
\label{reynolds}
\end{equation}
where $v$ has been arbitrarily chosen to be $10^{-2}$. 
In this sense the very early universe is almost a perfect conductor. Also,
the extremely large value of the Reynold's number indicates a turbulent
situation, which we shall find by other methods later.
\section{Aspects of relativistic MHD}
\subsection{Magnetohydrodynamics waves}

Let us begin by first presenting some general considerations.
In the framework of relativistic MHD in an expanding universe, we can still
discuss waves. Although the equations exhibited in the previous section are
considerably more complicated than their non-relativistic counterparts, the
MHD-waves are linear perturbations of the standard cosmological background.
Thus, the bulk velocity ${\bf v}$ must necessarily be small relative to the
velocity of light. It therefore follows that the displacement current can be
ignored. The background is homogeneous, and we assume the relativistic
relation $p=\rho$/3 for the background as well as for the fluctuations.
The continuity equation, i.e.\ (\ref{dlnrhodt1}), gives to the lowest
order the well known result $\rho={\rm const}$/$R^4$. To the next order we get
\begin{equation}
{\partial R^4 \delta \rho\over\partial t}+\frac{4}{3} R^4 \rho
\left(\frac{1}{R}\mbox{\boldmath $\nabla$}\cdot{\bf v}\right)=0,
\label{wave1}
\end{equation}
where $\delta \rho$ is the fluctuation in $\rho$. Also, from (\ref{Sexpression2})
we get to lowest order in the fluctuations
\begin{equation}
{\partial R^4\delta{\bf S}\over\partial t}
=-\frac{1}{R}~\mbox{\boldmath $\nabla$} R^4(\frac{1}{3}\delta \rho
+{\bf B} \delta {\bf B})+\frac{1}{R}
(R^2 {\bf B}) \mbox{\boldmath $\nabla$} (R^2 \delta {\bf B}).
\label{wave2}
\end{equation} 
Here $\delta{\bf S}=\frac{4}{3}\rho{\bf v}$, and $\delta {\bf B}$ is the fluctuation
of the background field ${\bf B}$, which is
assumed to behave like $\sim 1/R^2$. Of course, $\delta {\bf B}$ is expected
to have a similar scaling behavior as a function of time, but it also
has a spatial dependence.
Finally, we have the fluctuation equation
\begin{equation}
{\partial R^2\delta{\bf B}\over\partial t}=\frac{1}{R}\mbox{\boldmath $\nabla$}\times({\bf v}\times R^2{\bf B}),
\label{wave3}
\end{equation}
which follows from (\ref{induct1}), since the displacement current can be ignored
for small bulk velocities.

We now seek a wave solution which, because of the structure of
(\ref{wave1})-(\ref{wave3}), must contain the scale factor $R$ to the power $-2$, 
\begin{equation}
\delta {\bf B}=\frac{{\bf b}_0}{R^2}~\exp{i({\bf k} \cdot{\bf x}-\omega \tilde{t})}
\label{wave4}
\end{equation}
and
\begin{equation}
{\bf v}={\bf v}_0~\exp{i({\bf k}\cdot{\bf x}-\omega \tilde{t})},
\label{wave5b}
\end{equation}
\begin{equation}
\delta \rho=\frac{{\rm const.}}{R^4} \exp{i({\bf k} \cdot{\bf x}-\omega
\tilde{t})},
\label{wave5}
\end{equation}
where ${\bf b}_0$ and ${\bf v}_0$ are constants.
These expressions satisfy the basic fluctuation equations  
(\ref{wave1})-(\ref{wave3}) with
\begin{equation}
\tilde{t}=\int_{t_0}^{t} \frac{dt'}{R(t')},
\label{wavekuk}
\end{equation}
where $t_0$ is the initial time, in accordance with the results obtained in
the previous section.
 
We therefore see that with the scaling properties mentioned above the
equations are similar to the non-relativistic case (provided we use
the time $\tilde{t}$ in (\ref{wavekuk}). Thus we find the group velocity
$\partial \omega$/$\partial k=B$/$\sqrt{p+\rho}$. Because the
scaling properties of ${\bf B}$ and $\sqrt{p+\rho}$ with respect to the
expansion of the universe are the same, it follows that the group velocity is
independent of $R$. As far as the phase velocities are concerned, the
same is true. Assuming the background field to be in the $x$-direction,
then $\delta {\bf B}$ and ${\bf v}$ are in the $z$-direction, as in the
case of non-relativistic waves\cite{landau}.
One then finds that the velocities are given by
\begin{equation}
\frac{1}{2}\left(\sqrt{\frac{1}{3}+\frac{3{\bf B}^2}{16\pi\rho}+\frac{B_x}
{\sqrt{\pi\rho}}}\pm \sqrt{\frac{1}{3}+\frac{3{\bf B}^2}{16\pi\rho}
-\frac{B_x}{\sqrt{\pi\rho}}}~ \right).
\label{wave6}
\end{equation}
Of course, these velocities are given in terms of the conformal 
time $\tilde{t}$.
If we express the result in the standard time $t$, then (\ref{wave6}) should
be multiplied by 1/$R$.
It should be noted that the assumption of small bulk velocities can
only be maintained if $|{\bf B}|\ll \sqrt{\rho}$. If this condition is not
satisfied, we cannot expect the nonlinear effects to be small. 

\subsection{Two-dimensional slice}
Ideally, we would like to solve the MHD equations in three (plus one)
dimensions. However, as indicated in Sect.~II, this is a major computational
task. We restrict ourselves therefore to a two-dimensional slice only.
The main conclusion will be that much of the qualitative behavior of
nonrelativistic MHD carries over to the case of relativistic MHD.

We solve (\ref{tld2}) and (\ref{rhoexpression2})--(\ref{vxB}) numerically
using 6th order centered differences to compute the spatial derivatives
and a 3rd order Runge-Kutta scheme for the timestep.
We adopt random initial conditions for ${\bf B}$.
In order to guarantee that $\mbox{\boldmath $\nabla$}\cdot\tilde{\bf B}=0$ for all times, we
advance the $z$-component of the vector potential by means of the equation
$\partial\tilde A_z/\partial\tilde t={\bf e}_z\cdot({\bf v}\times\tilde{\bf B})$,
where $\tilde{\bf B}=\mbox{\boldmath $\nabla$}\times(\tilde A_z{\bf e}_z)$. The initial $A_z$ is
computed by solving $\nabla^2 A_z=-4\pi J_z$.
Initially ($t=t_0$, i.e. $\tilde t=0$) we put $\rho$ to unity.
Periodic boundary conditions are adopted in the $x$- and $y$-directions.

Our new equations (\ref{tld2}) + (\ref{rhoexpression2})--(\ref{vxB})
are scale-invariant, so it is sufficient to solve them on a computational
domain with size $L=1$. The results for a different domain size $L'$
are the same, but taken at a different time $\tilde t'=(L'/L)\tilde t$.

Like in all turbulence calculations there has to be some diffusion
to prevent the accumulation of energy at the smallest scale.
In order to restrict the effects of dissipation only to the largest
possible wave numbers we use hyperdiffusion, i.e. instead of the usual
diffusion operator $\nabla^2$, we use an operator of the form
$-\nabla^4$ for the evolution of all variables.
This technique is well known in turbulence research
(see e.g. \cite{Passot}). Also, since this procedure is merely of
computational relevance, we did not use the relativistic expressions.

The minimum diffusion coefficient $\nu$ we can afford is given by the 
empirical constraint that the ``mesh Reynolds number''
$Re_{\rm mesh}=U\delta x/\nu$ must not exceed the value $5-10$.
Here, $\delta x$ is the mesh size and $U$ is a typical velocity that
includes the velocity of waves and bulk motions. 
As was pointed out before, in the early universe the Reynolds
number is very large, which means that the magnetic diffusivity
$\eta=4\pi/\sigma$ should be much smaller than the adopted value of $\nu$.
In other words, in order to have realistic values of $\nu$,
$\delta x$ has to be extremely small.
However, the maximum number of meshpoints, $N=L/\delta x$, is limited
by computer memory and time. Our present, rather exploratory calculations
were carried out on a workstation, and so we restricted ourselves to
$N_{\max}=128$.
Even on larger computers we would never reach realistic values.
This demonstrates the difficulty of a realistic simulation.
It is obvious that numerical  simulations with a low Reynolds number 
cannot provide
a realistic picture of the early universe MHD. However, we believe they
are useful in illustrating the qualitative features of the problem.

\begin{figure}[htbp]
\epsfxsize=9.2cm\epsfbox{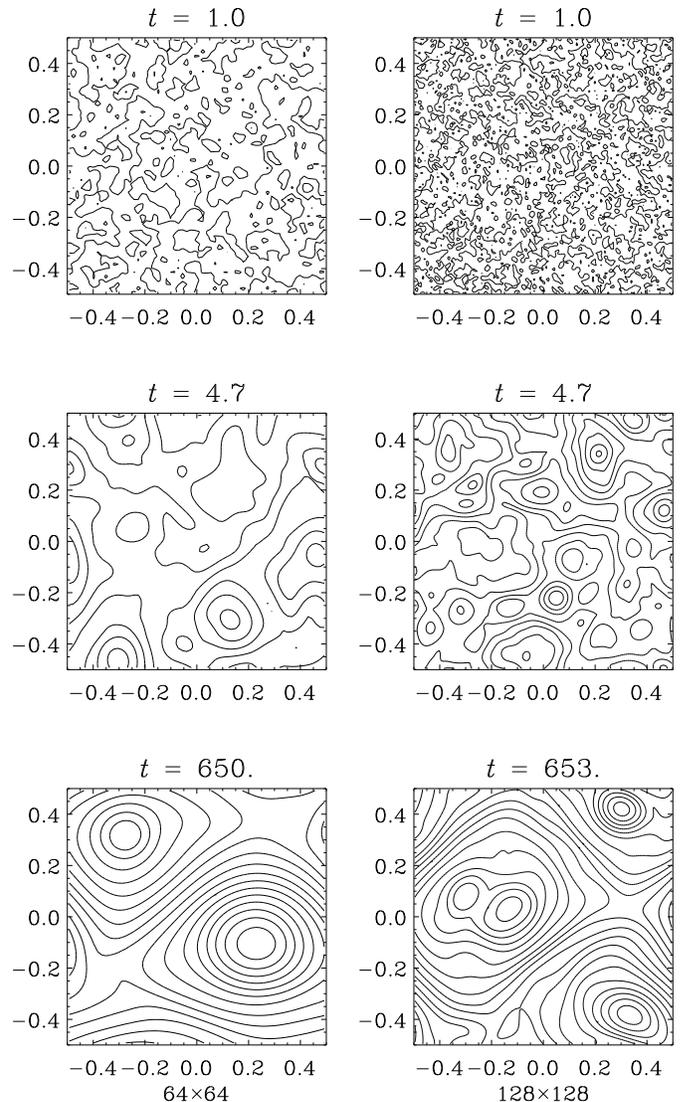}
\caption{
Left column: magnetic field lines at different times at low resolution
($64\times64$ meshpoints).
Right column: magnetic field lines at different times at higher resolution
($128\times128$ meshpoints).
}\label{ppsnap}\end{figure}

The evolution of the magnetic field is compared in Fig.~\ref{ppsnap}
for lower and higher resolution.
As time goes on, the coalescence of magnetic structures leads to the
gradual formation of larger and larger scales.
In the higher resolution case there are more small scale structures,
but also here the development of large scale fields is evident.
In turbulence research this phenomenon is known as an inverse cascade.
Such a cascade processes are linked to certain conservation properties
that the basic equations obey. For further details see ref.~\cite{inverse}.
We mention here only a few important aspects. An inverse cascade
exists both in two-dimensional and in three-dimensional MHD turbulence.
The only difference is that in the two-dimensional case it is an inverse
cascade of the magnetic potential, whereas in the three-dimensional case
it is an inverse cascade of the magnetic helicity density
${\bf A}\cdot{\bf B}$. In fact, the conserved quantities in the two cases
are $\int d^2x {\bf A}^2$ and $\int d^3x{\bf A}\cdot{\bf B}$, respectively.
For comparison we also mention that the difference between two and
three-dimensional hydrodynamic (non-magnetic) turbulence is more drastic.
In two-dimensional hydrodynamics there is an inverse energy cascade 
associated with the conservation of enstrophy (mean squared vorticity),
which has no counterpart in three-dimensional hydrodynamics.

The significance of an inverse cascade is that it leads to a transfer
of magnetic energy to larger and larger scales. This process is due to the
nonlinear terms giving rise to a mode interactions.
Energy spreads over different scales until some balance is achieved where
the kinetic and magnetic energy spectra have a certain slope.
In the ordinary MHD turbulence a relevant energy spectrum could be the
Iroshnikov-Kraichnan spectrum \cite{Kraichnan}, where the spectral energy
goes like $k^{-3/2}$, or a Kolmogorov type spectrum like $k^{-5/3}$.
These different spectra describe equilibrium situations, but in any case
it is clear that the spectrum will be very different from white noise,
which has a $k^{+2}$ power spectrum (see Sec.~IV.B below).
The possibility of energy transfer from small to large scales via an
inverse cascade could be of major importance in cosmology.
It could provide a seed field at the parsec or kiloparsec scale,
albeit at small amplitude.

\section{A cascade model}
\subsection{Description}

The ultimate goal is to solve the basic MHD equations in three dimensions
at high resolution, using random initial conditions.
Although we would be unable to cover a realistically large range of
length scales, it is important to know whether dynamo
action could be possible in a relativistic flow.
This is a major task, which would go beyond the scope of this paper.
To see the difficulties involved in such a program, the reader should
recall the difficulties in making long term weather predictions based
on the Navier-Stokes equations. Therefore,
in order to demonstrate some of the anticipated behavior of the full
1+3 MHD universe, we now study a cascade model of hydromagnetic turbulence.

In ordinary hydrodynamics and hydromagnetics many properties of
turbulence, in particular those related  to energy transfer and to the spectral
properties, including small intermittency corrections, have been studied
successfully using a simple cascade model \cite{cascade}.
This is true not only qualitatively, but also quantitatively, which is
the reason why the cascade model is now much used in studies of
nonlinear physics (see e.g.\ \cite{JPV91} and references therein).

The basic idea  is that the interactions due to the nonlinear terms
in the MHD equations
are local in wavenumber space. In $k$-space the quadratic nonlinear terms,
such as $\mbox{\boldmath $\nabla$}\times({\bf v}\times{\bf B})$, ${\bf v}\cdot\mbox{\boldmath $\nabla$}{\bf v}$, and
${\bf J}\times{\bf B}$,
become a convolution and have the general form \cite{GLPG}
\begin{equation}
{\bf N}_{\bf k}({\bf v},{\bf B})=
\int{\bf C}_{\alpha\beta}v_\alpha({\bf p}) B_\beta({\bf p}-{\bf k}) d^3 p~.
\end{equation}
(There are similar terms also for the other two nonlinearities.) where
${\bf C}_{\alpha\beta}={\bf C}_{\alpha\beta}({\bf k})$ is a tensor which is
linear in ${\bf k}$. Interactions in $k$-space involving triangles with similar
side lengths have the largest contribution, as discussed in \cite{GLPG}. 
This has led to the shell model (see e.g. \cite{JPV91} and references therein),
which is formulated in
the space of the modulus of the wave numbers. This space is approximated by
N shells, where each shell consists of wave numbers with $2^n\leq k \leq
2^{n+1}$ (in the appropriate units). The Fourier transform of the velocity
over a length scale $k_n^{-1}$ ($k_n=2^n$) is given by the complex quantity
$v_n$, and $B_n$ denotes a similar quantity for the $B$-field.
Furthermore, the convolution is approximated by a sum over the nearest and
the next nearest neighbors,
\begin{equation}
N_n(v,B)=\sum_{i,j=-2}^2 C_{ij} v_{n+i} B_{n+j}.
\end{equation}
Here $v$ and $B$ have lost their vectorial character, which
reflects the fact that this model is not supposed to be an approximation
of the original equations, but it should be considered as a toy model that
has similar $conservation$ properties  as the original equations. Thus e.g.
the energy flow should be represented by these equations.
It is quite remarkable that such models show several realistic features,
including intermittency corrections to the structure function exponents, and are
therefore rather popular both in the absence \cite{JPV91} and in the presence
\cite{Bis} of magnetic fields. Therefore we propose to apply such a
model also to the early universe.

Velocity and magnetic fields are thus represented by a scalar at the discrete
wave numbers $k_n=2^n$ ($n=1,...,N$), i.e. $k_n$ increases exponentially.
Therefore such a model can cover a large range of length scales
(typically up to ten orders of magnitude).
The important conserved quantity is $E_{\rm tot} R^4$, where
$E_{\rm tot}=\int T^{00}d^3x$ is the total energy.
Using that the bulk velocity is nonrelativistic, we have $\gamma\rightarrow1$,
so we can expand $\gamma^2\approx1+{\bf v}^2$. Hence
\begin{equation}
E_{\rm tot}\approx\int\left(\rho+{\textstyle{4\over3}}\rho{\bf v}^2
+{\textstyle{1\over2}}{\bf B}^2\right)d^3 x.
\label{Etot}
\end{equation}
Since we are here mostly interested in the evolution of the magnetic field
we ignore the detailed evolution of $\rho$ and assume
$\rho\approx\rho_0 R^{-4}$. Thus, we require that
\begin{equation}
\int\left({\textstyle{4\over3}}\rho_0{\bf v}^2
+{\textstyle{1\over2}}{\bf B}^2 R^4\right)d^3 x={\rm const}.
\end{equation}
We use ${\bf b}={\bf B} R^2$ and construct equations for $v_n$ and $b_n$ such that
\begin{equation}
{\textstyle{8\over3}}\rho_0 \sum_{n=1}^N v^*_n{dv_n\over d\tilde{t}}
+\sum_{n=1}^N b^*_n{db_n\over d\tilde{t}}=0.
\end{equation}
In computing the conservation of the energy, the complex conjugate of
this equation should be added. However, it turns out that the ``complex energy"
(exhibited in the above equation) is conserved by the following construction.

As pointed out, the main idea of the cascade model 
is to construct a set of equations that share the same basic
conservation properties of the nonlinear (quadratic) terms as the original
equations. Thus we write equations which mimic equations (\ref{tld2})
and (\ref{induct1}),
\begin{equation}
{\textstyle{4\over3}}\rho_0
{dv_n\over d\tilde{t}}=N_n(v,b),
\label{cascadeu}
\end{equation}
\begin{equation}
{db_n\over d\tilde{t}}=M_n(v,b),
\label{cascadeb}
\end{equation}
where
\begin{equation}
\begin{array}{lll}
2N_n(v,b)&=ik_n(A+C)
(v^*_{n+1}v^*_{n+2}-b^*_{n+1}b^*_{n+2})\\
                        &\!\!\!+ik_n(B-{\textstyle{1\over2}}C)
(v^*_{n-1}v^*_{n+1}-b^*_{n-1}b^*_{n+1})\\
                        &\!\!\!\!\!\!\!-ik_n({\textstyle{1\over2}}B+
{\textstyle{1\over4}}A)
(v^*_{n-2}v^*_{n-1}-b^*_{n-2}b^*_{n-1}),
\end{array}
\end{equation}
\begin{equation}
\begin{array}{lll}
M_n(v,b)&=ik_n(A-C)
(v^*_{n+1}b^*_{n+2}-b^*_{n+1}v^*_{n+2})\\
                        &\!\!\!+ik_n(B+{\textstyle{1\over2}}C)
(v^*_{n-1}b^*_{n+1}-b^*_{n-1}v^*_{n+1})\\
                        &\!\!\!\!\!\!\!-ik_n({\textstyle{1\over2}}B
                                            -{\textstyle{1\over4}}A)
(v^*_{n-2}b^*_{n-1}-b^*_{n-2}v^*_{n-1}),
\end{array}
\end{equation}
with $A$, $B$, and $C$ being free parameters.
It is straightforward to verify that $2\sum v^*_n N_n + \sum b^*_n M_n=0$,
using that $k_n=2^n$. The $\tilde{t}$ differentiations in Eqs.~(\ref{cascadeu})
and (\ref{cascadeb}) are included to mimic closely the nonrelativistic form of
eqs. (9) and (14). In the actual computations we have restored magnetic and
kinematic diffusion terms, $-\nu k_n^2 u_n$ and $-\eta k_n^2 b_n$, on the right 
hand sides of (\ref{cascadeu}) and (\ref{cascadeb}), respectively. 
We chose $\nu=\eta$ and as time goes on, lowered $\nu$ gradually  using the
formula $\nu\ge(\sum k_n^2 |u_n|^2)^{1/2}/k_{\max}^2$, where $k_{\max}=2^N$.
This formula estimates the minimum amount of diffusion necessary to prevent
the built-up of energy at the smallest resolved scale.
We use a 3rd order timestep, which is calculated via the formula
$\delta_t=0.25\min[(\sum k_n^2 |u_n|^2)^{-1/2}]$.

\subsection{Results}

\begin{figure}[htbp]
\epsfxsize=9cm\epsfbox{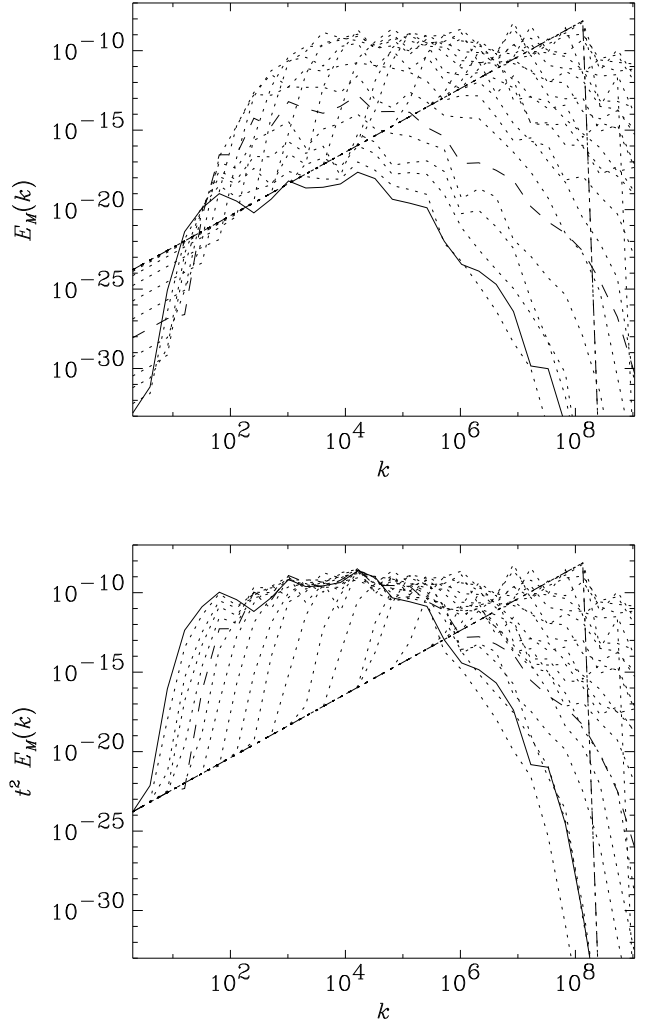}
\caption{
Spectra of the magnetic energy at different times.
The straight dotted-dashed line gives the initial condition ($t_0=1$),
the solid line gives the final time ($t=3\times10^4$), and the dotted
curves are for intermediate times (in uniform intervals of 
$\Delta\log(t-t_0)=0.6)$. $A=1$, $B=-1/2$, and $C=0$.
}\label{pb_array}\end{figure}

The numerical study of the cascade model requires of course that the
parameters $A,B,C$ are fixed. This problem turns out to be quite
interesting, since it allows one to associate the cascade model with a
dimension. In hydrodynamics the parameters are fixed by taking into account  
a conservation law which is non-trivial in the dimension considered. In
two dimensions, for example, one uses the requirement that the enstrophy is conserved,
whereas in three dimensions the helicity should be conserved.
In three dimensions Jensen et al. \cite{JPV91} used the values $A=1$,
$B=-1/2$, and $C=0$, which we have adopted also in several models presented here.
We compare the results with another set of parameter for which the quantity
\begin{equation}
H_M=\sum(-1)^n k_n^{-1} b^*_n b_n,
\label{helicity_shell}
\end{equation}
is conserved, in addition to the total energy \cite{Bis}.
This requires that $A=7/5$, $B=-1/10$, and $C=1$.
The quantity $H_M$ resembles the magnetic helicity ${\bf A}\cdot{\bf B}$,
which is important, because associated with it is the inverse
cascade of magnetic helicity and energy \cite{PFL76}.
In Fig.~\ref{pb_array} we plot the spectral magnetic energy density
$E_M(k_n)=|b_n|^2/(k R^4)$ computed for these values, with the initial
field taken to be random (i.e., $E_M(k)=k^2$). The reason we interpret this
expression as the magnetic energy is that we know that $\sum b^*_n b_n$
enters in the conserved energy. However, $\sum\sim\int dn=\int dk$/$(k$ log 2),
so $E_M(k_n)$ is the energy in $k$-space.
We used $N=30$, which covers a range of length scales of
approximately ten orders of magnitude. As one can clearly see, magnetic
energy is transferred from small scales to large scales. This is called the
inverse cascade effect. Such an effect is found in many nonlinear systems,
for example in two dimensional turbulence, relevant e.g. for the atmosphere.

\begin{figure}[htbp]
\epsfxsize=8.2cm\epsfbox{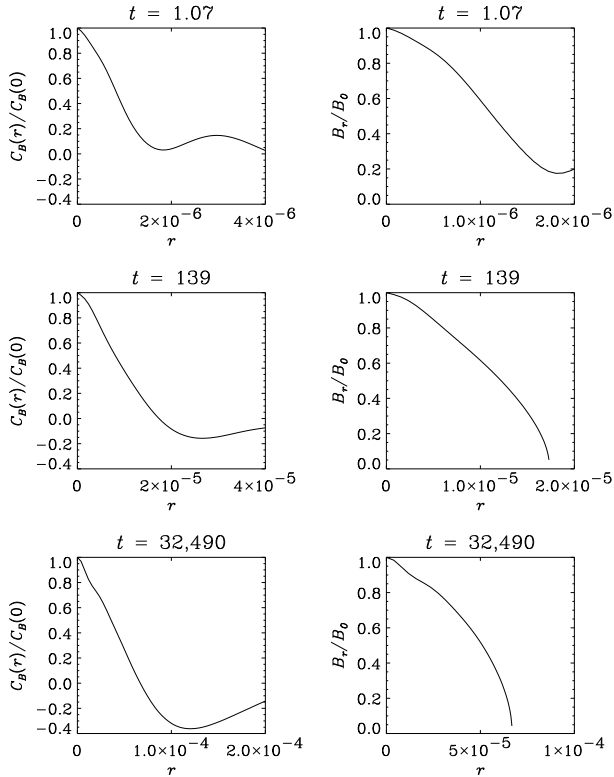}
\caption{Left column: The correlation function of $B$ for three different
times. Right column: The RMS magnetic field as a function of distance
for three different times.}\label{pcorrel3}\end{figure}

The quantity of paramount interest is the the magnetic field correlation
function
\begin{equation}
C_B(r)\equiv\langle B(r+x) B(x)\rangle,
\end{equation}
which is related to the power spectrum via a Fourier transform,
$C_B(r)=\int E_M(k)\cos(kr) dk$. It is difficult in general to compute this
quantity, due to the fluctuations in the spectrum $E_M(k)$. Therefore we have  
computed $C_B(r)$ from the spectra of the cascade model by interpolating
$E_M(k)$ on a uniformly spaced mesh. This of course introduces 
some uncertainty. The result is shown in Fig.~\ref{pcorrel3}.
Note the clear increase of the widths of the correlation functions.

Note also the anticorrelation at larger length scales. For a magnetic field
this is natural, because if one considers the field in some
region from a point far away from this region, the magnetic field in the region
appears to be approximately a dipole. A negative correlation then arises
because the field loop has to close. This would then basically be a consequence
of div ${\bf B}$=0, which has a tendency to lead to negative correlations.
However, the cascade model of course has the difficulty that it does
not really operate in ordinary space, but instead it is formulated in the
modulus of ${\bf k}$-space. Hence we cannot really investigate to what extent
the condition div ${\bf B}$=0 is satisfied, in contrast to the two dimensional
case discussed in Sec.~III.B.

Another quantity of interest is the average magnetic field as a function of
distance,
\begin{equation}
{\bf B}(r)\equiv \frac{1}{r^D}\int d^D x {\bf B}({\bf x}),
\label{afstand1}
\end{equation}
where the integration is over a volume of size $r^D$ in $D$-dimensions. From
this definition we have
\begin{equation}
\langle{\bf B}(r)^2\rangle=\frac{1}{r^{2D}} \int d^D x \int d^D y
\langle{\bf B}({\bf x}) {\bf B}({\bf y})\rangle,
\label{afstand2}
\end{equation}
where both integrations are over a volume of size $r^D$.
Thus the root mean square magnetic field
\begin{equation}
B_r=\langle{\bf B}(r)^2\rangle^{\frac{1}{2}}
\label{afstand3}
\end{equation}
can be computed directly from the correlation function $C_B(r)$ via
\begin{equation}
B_r=\left({1\over r^3}\int_0^r r'^2 dr' C_B(r')\right)^{1/2}
\label{Br_CB}
\end{equation}
For a random field, $B_r$ behaves like $r^{-\frac{D}{2}}$, so the
interesting question is whether this initial behavior changes as time
passes. In Fig.~\ref{pcorrel3} we show the results. There is a clear
broadening of $B_r$ towards larger distances as time passes, as we
would expect from the inverse cascade behavior.

\begin{figure}[htbp]
\epsfxsize=9cm\epsfbox{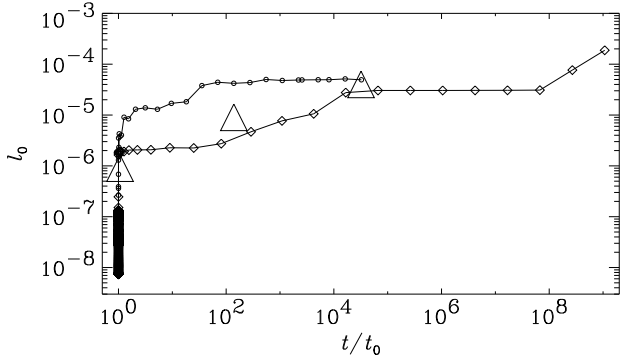}
\caption{
Evolution of the integral scale. The plot symbols denote uniform intervals of
$\Delta\log(t-t_0)=0.6$. The circular points corresponds to the hydrodynamic
values $A=1$, $B=-1/2$, and $C=0$, the diamonds correspond to the MHD values
$A=7/5$, $B=-1/10$, and $C=1$, and the triangles are the widths computed
for the correlation function.}\label{pb_array_intscl}\end{figure}

The determination of the width of the correlation function above is not
very accurate because of the fit involved in computing the Fourier
transform of the spectrum.  We shall therefore now introduce another
length scale that is easier to compute, but whose value is similar to
the width of the correlation function. The relevant length scale in
turbulence theory is the so-called integral scale, which is the
characteristic length associated with the large energetic eddies of
turbulence. Roughly speaking one could view it as a measure of the
coherence of the magnetic field, too. It is defined by
\begin{equation}
l_0=\left.\int 2\pi k^{-1} E_M(k) dk\right/\int E_M(k) dk,
\end{equation}
which, in our cascade model, corresponds to
$l_0=\sum 2\pi k_n^{-1}|b_n|^2/\sum |b_n|^2$.
If the spectrum is random we get $l_0\simeq{3\over2}2\pi k_{\max}^{-1}$, where
$2\pi/k_{\max}$ is the shortest length scale present in the model.
This length scale in the initial random spectrum is determined by the
mechanism generating the primordial field.
In Fig.~\ref{pb_array_intscl} we show the evolution of $l_0(t)$ in two cases,
namely for the hydromagnetic $A,B,C$ (circular points) and for the MHD $A,B,C$
(diamond-shaped points). Although the two sets of values for $A,B,C$ do not
yield identical results, we see that the curves are qualitatively similar.
In both cases $l_0$ increases rapidly by 4-5 orders of magnitude, and 
there is a plateau structure. The MHD result (diamond-shaped points) has
a plateau stretching to $t/t_0=10^8$, but for larger times $l_0$ keeps
increasing. The increase of $l_0$ by almost 5 orders of magnitude is
important, because it could lead to magnetic fields at the present time at
length scales comparable to one parsec. If we take the electroweak phase
transition as the initial state,  the QCD phase transition
occurs approximately for $t$/$t_0$=$10^6$. The maximum time $t$/$t_0$=
$10^9$ reached in our simulation corresponds to a temperature of 3 MeV,
which is close to the nucleosynthesis.
It should be noticed that from these results one cannot, of course,
say anything about what happens at later times. Therefore it could be that
$l_0(t)$ increases further, either by reaching new plateau(s), or otherwise.

We also measured the integral scale of the magnetic field in the 
two-dimensional model of Sec.~III.B and found a clear increase with time.
For the three times plotted in Fig.~1 we found for the $64\times64$ case the
values $l_0=0.09$, 0.50, and 0.95, whereas in the $128\times128$ case
values were $l_0=0.04$, 0.28, and 0.76. The initial difference of a factor
two is due to the different resolution. At later times the integral scales
for low and high resolution are more similar.

The evolution of $l_0$ is not straight, but if we make a linear fit
through the values given by the diamonds in Fig.~\ref{pb_array_intscl}
we find (ignoring the steep initial increase)
\begin{equation}
l_0(t)\approx r_0 (t/t_0)^{1/4}.
\label{linie2}
\end{equation}
For further applications of the rough fit formula
(\ref{linie2}) it may be more sensible to express time in terms of
temperature, $T\propto t^{-1/2}$, so
\begin{equation}
l_0(T)\approx r_0 (T/T_0)^{-1/2},
\label{linie3}
\end{equation}
where $r_0\approx10^{-6}$; see Fig.~\ref{pb_array_intscl}.
If we were to extrapolate to the present time we first have to fix the
scale $r_0$ by physical arguments. The various models presented in the
literature \cite{pt,ven} give characteristic scales for the primordial
field when it is generated. This scale should be identified with the
lowest scale in our calculations which, in the case of
the shell model, is about $10^{-8}$.
The scale $r_0$ is typically somewhat larger ($10^{-6}$ in the shell
model). The reason for this is presumably that a purely random initial
condition is not consistent with the MHD equations.

In order to clarify these point we take an example. If we assume that
at the time of the electroweak phase transition ($T=100\,$GeV)
$r_0$ was $10^{-3}\,$cm (the horizon scale was $\approx4\,$cm) then,
using our extrapolation (\ref{linie3}), we arrive at a scale of 2 parsec.
If we assume that the initial magnetic field was $10^{18}\,$G, then
the present day value would be $10^{-11}\,$G.
Such values would lead to sufficiently strong seed magnetic 
fields to explain the field even in high redshift galaxies by
dynamo action \cite{fields}.
This extrapolation may be too naive, because the nature of turbulence
will change as the universe cools down. Furthermore, at later times,
when structure formation begins, gravitational energy may lead to additional
stirring and enhancement of turbulence in localized regions.

\section{Discussion}

In the two-dimensional case we found that, starting from a small scale magnetic
field, magnetic structures develop at progressively larger scales.
This process of self-organization corresponds to an inverse cascade of
magnetic energy and helicity.
Using then a cascade (or shell) model to study three-dimensional MHD turbulence
we were able to follow this inverse cascade over much longer times.
Such a cascade model has been rather successful in the study of
hydrodynamic turbulence.

The possibility of an inverse cascade means that the scale of fluctuations of
the primordial magnetic field increases much beyond its original scale
given by particle physics.
Taking the parameter $l_0$ as a measure of the coherence length of the
magnetic field, we see that there is an increase in the coherence of 4 to 5
orders of magnitude. This means that previous estimates of the field strengths
in various mechanisms for generating a primordial field should be revised
accordingly. For example, let us consider the estimate by Vachaspati in
ref. \cite{pt}. Taking the area average one has the estimate
\begin{equation}
B_r\sim gT^2/4N,
\label{vas1}
\end{equation}
where $N$ is the number of steps needed to reach a given scale in terms of the
``fundamental" scale at which the field is generated. In this case the
fundamental scale is the electroweak scale \cite{pt}. Proceeding as in ref.
\cite{pt} one has $N\sim 10^{24}$ today, if the relevant scale is of order
100 kpc. However, due to the MHD corrections $N$ is, from a conservative
point of view, reduced because of the effect of turbulence, and one would
instead have $N\leq 10^{19}$, which reduces the stochastic decrease of $B_r$.
It should be emphasized that from our calculations one can only say
what happens up to a time of order $10^9~ t_{\rm EW}$, so presumably
$N$ is considerably below $10^{19}$ today. 

The turbulent nature of the magnetic field may have interesting effects on
the various phase transitions in the early universe. Also, the
inherent shift of energy from small to large scales may be of interest in
connection with the density fluctuations due to the magnetic energy.   

Of course, the cascade model is a {\it model} of the real 1+3 dimensional
MHD turbulence. Its successful application in many, widely different
nonlinear physical problems suggests, however, that it might also
be applicable to the primordial magnetic fields of the early universe.
Therefore we believe that its indication of the strong increase
in the coherence scale of the primordial  field should be taken
seriously, and that further, more detailed studies are warranted.

\acknowledgments
We thank Mogens H\o gh Jensen for interesting discussions.

\appendix
\section{}
We give here the equations for the case where the bulk velocity is small.
(The gas remains still relativistic, which is important for the
scaling properties of $\tilde\rho$ and $\tilde p$.)
\begin{equation}
{\partial\ln\tilde\rho\over\partial\tilde t}
=-{4\over3}({\bf v}\cdot\mbox{\boldmath $\nabla$}\ln\tilde\rho
+\mbox{\boldmath $\nabla$}\cdot{\bf v})
-{\tilde{\bf J}\cdot\tilde{\bf E}\over\tilde\rho},
\end{equation}
\begin{eqnarray}
{D{\bf v}\over D\tilde t}=
-{\bf v}\left({D\ln\tilde\rho\over D\tilde t}
+\mbox{\boldmath $\nabla$}\cdot{\bf v}\right)
-{1\over4}\mbox{\boldmath $\nabla$}\ln\tilde\rho
+{\tilde{\bf J}\times\tilde{\bf B}\over{4\over3}\tilde\rho},
\label{vel}
\end{eqnarray}
where $D/D\tilde t=\partial/\partial\tilde t
+{\bf v}\cdot\mbox{\boldmath $\nabla$}$
is the total derivative, and
\begin{equation}
{\partial\tilde{\bf B}\over\partial\tilde t}=\nabla\times({\bf v}\times\tilde{\bf B}),
\quad\tilde{\bf J}={1\over4\pi}\mbox{\boldmath $\nabla$}\times\tilde{\bf B}.
\label{in}
\end{equation}



\begin{thebibliography}{X}
\bibitem{frozenin}
It is often said that since the magnetic field is ``frozen in" at large
scales, it must decrease exactly like 1/$R(t)^2$ according to MHD.
However, the charged plasma can have a bulk velocity (peculiar
velocity) due to its interaction with the magnetic field, and this may
lead to shifts of the coherence scale of the field, and hence the 1/$R^2$
decrease does not necessarily give a full description of the situation.
\end{thebibliography}
\end{document}